\documentclass[12pt]{article}
\usepackage{epsf}

\begin{document}

\title{Induced parity violating thermal effective action for
$2+1$-dimensional fermions interacting with a non-Abelian background}

\author{F. T. Brandt$^a$, Ashok Das$^b$ and J. Frenkel$^a$
\\ \\
$^a$Instituto de F\'{\i}sica,
Universidade de S\~ao Paulo\\
S\~ao Paulo, SP 05315-970, BRAZIL\\
$^b$Department of Physics and Astronomy\\
University of Rochester\\
Rochester, NY 14627-0171, USA}
\date{\today}
\maketitle

\begin{abstract}
We study the parity breaking effective action in $2+1$ dimensions,
generated, at finite temperature, by massive fermions interacting with
a non-Abelian gauge background. We explicitly calculate, in the static
limit, parity violating amplitudes up to the seven point
function, which allows us to determine the corresponding effective
actions. There are two classes of such actions that arise, namely,
terms that do not manifestly depend on $\vec{E}$ and ones that do. We
derive the exact effective action that is not manifestly dependent on
$\vec{E}$. For the other class that depends explicitly on
$\vec{E}$,  there are families of terms that
can be determined order by order in perturbation theory. We attempt to
generalize
our results to non-static backgrounds through the use of time ordered
exponentials and prove gauge invariance, both {\it small} and {\it
  large}, of the resulting effective action. We also point out some open
questions that need to be further understood.
\end{abstract}
\newpage
\section{Introduction:}
Chern-Simons theories in $2+1$ dimensions
\cite{chern:1974,Deser:1982wh} have  attracted a lot of
attention  in the past few years for a variety of reasons 
\cite{dunne:leshouches}.
One of the issues  studied extensively, for example, is the question
of  {\it  large}  gauge invariance at finite temperature
\cite{babu:1987rs}-\cite{cdfosco}.
In particular, it is known for sometime now that, at
finite temperature, massive fermions interacting with a background
non-Abelian gauge field induce a Chern-Simons term with a coefficient
which is a continuous function of temperature \cite{babu:1987rs}.
Therefore, at an
arbitrary temperature, the Chern-Simons coefficient cannot have a
discrete value as would be necessary for invariance under {\it large}
gauge transformations.

There has been a lot of progress in understanding this puzzle in the
past few years. It has been shown, for example, within the context of
$0+1$ dimensional fermions interacting with an external Abelian gauge
field, that at finite temperature, an infinity of terms is induced in
the effective action \cite{dunne:1997yb} in such a way that {\it
  large}  gauge invariance is
restored in the complete effective action, even though at any finite
order in perturbation theory such an invariance will not be
manifest \cite{dunne:1997yb,das:1998gg}.
Subsequently, this analysis has been generalized to $2+1$
dimensional fermions interacting with an Abelian gauge
background \cite{deser:1997nv}-\cite{fosco:1997vu}.
Although, at zero temperature, Abelian gauge
transformations correspond to transformations with a trivial winding
number and impose no restrictions on the Chern-Simons coefficient, at
finite  temperature, because of periodicity (in the imaginary time
formalism), Abelian transformations with nontrivial winding are
possible. Furthermore, at finite temperature, amplitudes and,
therefore, the effective action become non-analytic functions at the
origin in the energy-momentum plane \cite{weldon:1993bv}-\cite{das:book97}
(because of additional channels
available for processes in a thermal medium). In the case of Abelian
gauge backgrounds, we
understand  the resolution of the puzzle of {\it large} gauge
invariance, at least, in two distinct conventional limits. In the long
wave (LW) limit, where all spatial momenta vanish, it has been shown
through explicit perturbative calculations up to the four point
function that  {\it large} gauge invariance is not a problem order by
order in perturbation theory \cite{Brandt:2000dd}.  On the other
hand, in the static limit, where all energies vanish, at any order in
perturbation, {\it large} gauge invariance is not manifest. However,
it is also known that one can sum the leading order terms in the
parity violating effective action in this
limit and the resulting effective action has a form
which is a generalization of the $0+1$ dimensional result and is
invariant under {\it large} gauge transformations much the same
way \cite{deser:1997nv}-\cite{fosco:1997vu}.
The leading order parity violating effective action, in this limit, also
corresponds to the exact effective action (parity violating) when the
electric field vanishes. Beyond the leading order, one picks up contributions which
are proportional to the electric field and all such terms are
manifestly {\it large} gauge invariant \cite{Brandt:2001jc}. 
It is worth pointing out here that, for any intermediate limit 
(between the two conventional limits, namely, LW and static), 
calculations become extremely complicated owing to the analytic 
continuation that is needed in the imaginary
time formalism. As a consequence, explicit forms for the parity
violating effective action are not available in this regime. However,
it is natural to believe that {\it large} gauge invariance will hold
in the complete effective action in this case as well.

The studies in the Abelian gauge background have given us very
valuable insights into the question of {\it large} gauge invariance in
such theories. With this knowledge, therefore, it is natural to
reanalyze the question of {\it large} gauge invariance for a $2+1$
dimensional massive fermion interacting with a non-Abelian gauge background at
finite temperature (which, in fact, led to all these studies). We
would like to point out that such an analysis was, in fact, carried
out earlier \cite{fosco:1997vu}, in the path integral formalism, for
a very restrictive
gauge background of the form (gauge potentials are matrices 
belonging to $SU(N)$)
\begin{eqnarray}
A_{i} = A_{i}(\vec{x}),\qquad A_{0} \;\;\;= & A_{0}(t)\nonumber\\
\left[A_{0}, A_{i}\right] \;\;\;= & \!\!\!\!\!\!\!\!\!\! 0\label{1}
\end{eqnarray}
Such a background corresponds to a vanishing electric field and it was
shown  that the resulting form of the parity violating
effective action, in this background, had an Abelian structure, which
was  a simple generalization  of that for the Abelian case.
However, it is generally believed that the Abelian form of this
parity violating effective action is a consequence of the above restriction
on the background fields. 

To analyze the true non-Abelian structure of the parity violating
effective action at finite temperature, we have chosen a more direct
approach. Namely, we study the amplitudes in perturbation theory, at
finite temperature, in the static limit. We choose the static limit
for two reasons: i) it is in this limit that {\it large} gauge
invariance is not manifest in perturbation theory in the Abelian case
and, ii) calculations are much more complicated in other limits. We
have calculated amplitudes up to the seven point function and our
conclusions are as follows. We find that the effective action has two
classes of terms - one that manifestly depends on $\vec{E}$ and another
which does not. In the static limit, we find that the parity violating
effective action, which does not manifestly depend on $\vec{E}$  has a
form  similar to that in \cite{fosco:1997vu}
(with proper non-Abelian terms). For a vanishing electric field, the
non-Abelian structures drop out and our action coincides with that
obtained in \cite{fosco:1997vu}, although our gauge background is more general.
We give a path integral derivation of this, for our choice of
backgrounds,  along the lines of
\cite{fosco:1997vu} showing that this is indeed the
exact parity 
violating effective action in the static limit, when the electric
field  vanishes. However, unlike the Abelian case, it is not true that 
this is the leading order effective action in the static limit. 
The static parity violating effective action, which manifestly depends
on $\vec{E}$, has a family of other terms, which can also 
contribute at the leading order. Some of these additional terms, in
fact, do contribute non trivially at zero
temperature and, therefore, can be given a Lorentz invariant
description. 
This is, in fact, completely consistent with
the non-Abelian Ward identities of the theory (namely, in the Abelian
theory, the Ward identities would imply that the $n$-point function
with $n>2$ is, at least, of the order of $p^{n}$
\cite{coleman:1985zi}; however,   non-Abelian
Ward identities do allow dependence on momenta of lower order).

The parity violating static effective action is manifestly invariant
under residual static non-Abelian gauge transformations, which are
{\it small} gauge transformations. In fact, in the strict static
limit, there can be no {\it large} gauge transformation and,
therefore, to analyze the question of {\it large} gauge invariance,
one has to go away from the static limit. As we have already
argued, such a calculation is extremely difficult and,
consequently, based on the results of our calculations, we have tried
to look for possible generalizations of our results away from the
strictly static limit. We derive a possible form of the parity
violating effective action, drawing from the studies in the Abelian
theory,  that will be both {\it small} and {\it large} gauge
invariant. The situation, however, is very different from the Abelian
case and it seems to us that many issues remain to be clarified before
we understand fully the question of {\it  large}  gauge invariance in
such a case. 

The paper is organized as follows. In section {\bf 2}, we present our
results for the amplitudes, up to the seven point function, in the
static limit for terms which do not manifestly depend on $\vec{E}$. We
show  that these
amplitudes can be derived from an action, whose form is similar to that
of the parity violating effective action in \cite{fosco:1997vu},
but with genuine non-Abelian structures. We show that when the electric field
vanishes, the non-Abelian interaction terms indeed drop out and our
action coincides  with that in
\cite{fosco:1997vu}.  We give an
alternate  path integral derivation,
showing that this parity violating effective action is exact for the 
case when the electric field is trivial in this limit. In
section {\bf 3}, we present the amplitudes for the terms which
manifestly depend on $\vec{E}$. These amplitudes satisfy the
non-Abelian Ward identity following from the residual static gauge
invariance of the theory. It is shown that some of these amplitudes do
contribute at the same leading order as those in section {\bf 2} and,
therefore, unlike the Abelian case, the effective action that does not
manifestly depend on $\vec{E}$, cannot be thought of as the leading
order  term in 
the static limit. We present an effective action that would generate 
these amplitudes. There are two classes of terms in this effective
action, i) terms with a nontrivial limit at zero temperature have a 
covariant form and, ii) those with a vanishing limit at zero 
temperature have a non-covariant structure. In section {\bf 4},
we try to generalize our results away from the strictly static limit.
Following closely to the derivation in the Abelian case, we propose a
possible generalization of the parity 
violating effective action to non-static backgrounds that will be
both {\it small} and {\it large} gauge invariant. We also discuss
various issues that remain to be clarified and present a brief
conclusion in section {\bf 5}. In 
appendix {\bf A}, we compile some useful finite temperature relations.
In appendix {\bf B}, we discuss general
properties of thermal gauge transformations and the consequences of a
vanishing electric field, both in the Abelian as well as the
non-Abelian theory. In appendix {\bf C}, we discuss briefly,
how the restriction in Eq. (\ref{1}) necessarily leads to an Abelian
structure for the parity violating effective action.

\section{Manifestly $\vec{E}$ independent parity violating effective
action:} 

We are considering $2+1$ dimensional massive fermions interacting with a
non-Abelian gauge background described by the Lagrangian density
\begin{equation}
{\cal L} = \overline{\psi}\left(i\gamma^{\mu}D_{\mu} - M\right)\psi
\end{equation}
where we assume $M>0$ for simplicity. The fermions are assumed to belong to the
fundamental representation of $SU(N)$ so that the covariant derivative is
defined to be
\[
D_{\mu} = \partial_{\mu} + ig A_{\mu}
\]
where the gauge fields, $A_{\mu}$, are matrices belonging to the
fundamental representation of $SU(N)$.

In calculating the amplitudes at finite temperature, we will use the
imaginary time formalism, where the time axis is rotated to a finite
interval in the imaginary axis (we refer the readers 
to \cite{das:book97,kapusta:book89,lebellac:book96} for details). In
this  Euclidean space,  the three Dirac matrices can be
chosen to be anti-Hermitian and a particular representation can be
chosen to be (although our results are independent of any choice of
the representation)
\[
\gamma_{0} = i\sigma_{3},\quad \gamma_{1} = i\sigma_{1},\quad
\gamma_{2} = i\sigma_{2}
\]
We are interested in calculating the amplitudes in the static limit,
which corresponds to a choice of the background fields of the form
\begin{equation}
A_{0} = A_{0}(\vec{x}),\qquad A_{i} = A_{i}(\vec{x})\label{2}
\end{equation}
without any further restriction on $\vec{E}$.

In this static background, the action has a residual gauge invariance
of the form  
\begin{eqnarray}
\psi & \rightarrow & U^{-1}(\vec{x})\psi\nonumber\\
A_{0} & \rightarrow & U^{-1}(\vec{x})A_{0}U(\vec{x})\nonumber\\
A_{i} & \rightarrow & U^{-1}(\vec{x})A_{i}U(\vec{x}) - {i\over g}
U^{-1}(\vec{x})\partial_{i}U(\vec{x})\label{2a} 
\end{eqnarray}
As a consequence of this symmetry, it is straightforward to derive
that the gauge amplitudes will have to satisfy the Ward identities
following from
\begin{equation}
\partial_{i}{\delta\Gamma_{\rm eff}\over \delta A_{i}^{a}} + gf^{abc}
A_{0}^{b} {\delta\Gamma_{\rm eff}\over \delta A_{0}^{c}} + gf^{abc}
A_{i}^{b} {\delta\Gamma_{\rm eff}\over \delta A_{i}^{c}} = 0\label{2b}
\end{equation}
where $\Gamma_{\rm eff}$ is the effective action resulting from evaluating
the fermion loops.

As we have already mentioned in the introduction, our calculations of
the amplitudes give rise to two classes of effective actions - one
that  manifestly depends on $\vec{E}$ and another which does not. In
this section, we will concern ourselves only with the class that does
not depend manifestly on $\vec{E}$. From the structure of the terms
which depend manifestly on $\vec{E}$ (which we will discuss in the
next section), it is easy to recognize that these may be thought of as
resulting from the full effective action when the electric field is
covariantly constant. Although the vanishing electric field is a
subclass of these configurations, we do not, in fact, assume the
electric field to vanish and we will comment more extensively on this
special subclass later in this section.

The calculation of the amplitudes is tedious, but straightforward, 
and we will not give details of the calculation which
have been described earlier \cite{Brandt:2000dd, Brandt:2001jc}
(within the  context of an Abelian
background). However, let us define some notation to
present the results of our calculations in a more manageable form. Let
\begin{equation}\label{Isum}
I^{(r+1)} = {M\over 4\pi r\beta} \sum_{n} {1\over
  (M^{2}+\omega_{n}^{2})^{r}}\label{4}
\end{equation}
where
\begin{equation}
\omega_{n} = {(2n+1)\pi\over \beta}
\end{equation}
represents the Matsubara frequencies for fermions and $\beta = {1\over
  T}$ with $T$ representing the temperature. These quantities
can all be evaluated in a closed form by successive differentiation of
(see appendix {\bf A} for the explicit forms of some of the lower order
$I^{(r)}$'s)
\begin{equation}
I^{(2)} = {1\over 8\pi} \tanh {\beta M\over 2}
\end{equation}
Let us also define a completely symmetrized fourth rank tensor in the
internal space of the form
\begin{equation}
\Delta^{abcd} = \delta^{ab}\delta^{cd} + \delta^{ac}\delta^{bd} +
\delta^{ad}\delta^{bc}\label{4a}
\end{equation}

All the amplitudes can be represented in terms of these quantities in
the following way. First, let us  note some essential
features of the amplitudes in the static limit. In this limit, the
parity violating amplitudes involve only an odd number of $A_{0}$
fields. The amplitudes, which lead to the effective action that does
not manifestly depend on $\vec{E}$, have the
following forms ($\vec{k}_{1},\vec{k}_{2},\cdots$ correspond to the
external momenta associated with the first, second,$\cdots$ indices,
all incoming; for the two point function, the momentum is associated
with the second index)
\begin{eqnarray}
\Pi_{0i}^{ab, (1)} & = & -
g^{2}\delta^{ab}\epsilon_{ij} k_{j}\,I^{(2)}\nonumber\\
\Pi_{0ij}^{abc, (1)} & = & ig^{3}f^{abc}\,I^{(2)}\epsilon_{ij}\nonumber\\
\Pi_{000i}^{abcd, (1)} & = & g^{4}\epsilon_{ij}k_{4j}\left({2\over
    N} \Delta^{abcd} + 
d^{abe}d^{cde}+d^{ace}d^{bde}+d^{ade}d^{bce}\right)\nonumber\\
 &  &\;\;\; \times \left(I^{(3)}-2M^{2}I^{(4)}\right)\label{5}
\end{eqnarray}
Here, $f^{abc}$ and $d^{abc}$ denote respectively the anti-symmetric
and symmetric structure constants for $SU(N)$.

As we go to higher point amplitudes, the calculation involves a color
trace over more and more color matrices and as a result, the color
factors become more and more complicated. Therefore, we will present
the results of our calculations of higher point functions only for
$SU(2)$, where we obtain
\begin{eqnarray}
\Pi_{000ij}^{abcde, (1)} & = & - ig^{5}\epsilon_{ij}
(\delta^{ab}\epsilon^{cde}+\delta^{ac}\epsilon^{bde}+
\delta^{bc}\epsilon^{ade}) 
\left(I^{(3)}-2M^{2}I^{(4)}\right)\nonumber\\
\Pi_{00000i}^{abcdef, (1)} & = &  {g^{6}\over 2}
\epsilon_{ij}k_{6j} C^{abcdef}
\left(3I^{(4)}-16M^{2}I^{(5)}+16M^{4}I^{(6)}\right)\nonumber\\
\Pi_{00000ij}^{abcdefg, (1)} & = & -{ig^{7}\over 2}
\epsilon_{ij} C^{abcdefg}
\left(3I^{(4)}-16M^{2}I^{(5)}+16M^{4}I^{(6)}\right)\label{6}
\end{eqnarray}
where
\begin{eqnarray}
C^{abcdef} & = & \sum_{i=1}^{5} C_{i}^{abcdef}\nonumber\\
C_{1}^{abcdef} & = &
\delta^{ef} \Delta^{abcd}, \quad C_{2}^{abcdef} =
\delta^{df} \Delta^{abce}\nonumber\\
C_{3}^{abcdef} & = & \delta^{cf} \Delta^{abde}, \quad C_{4}^{abcdef} =
\delta^{bf} \Delta^{acde},\quad C_{5}^{abcdef} = \delta^{af}
\Delta^{bcde}\nonumber\\ 
C^{abcdefg} & = &
\epsilon^{afg} \Delta^{bcde} + 
\epsilon^{bfg} \Delta^{acde}
  + \epsilon^{cfg} \Delta^{abde} + \epsilon^{dfg} \Delta^{abce}
 + \epsilon^{efg} \Delta^{abcd}\label{7}
\end{eqnarray}

From the definition in Eq. (\ref{4}), it is easy to see that the
parity violating amplitudes, to this order, completely
coincide with those following from the action
\begin{equation}
\Gamma_{\rm eff}^{{\rm PV}, (1)} = {ig\over 2\pi} \int d^{2}x\,{\rm
  Tr}\,\arctan\left(\tanh {\beta M\over 2}\,\tan {g\beta A_{0}(\vec{x})\over
  2}\right) B(\vec{x})\label{8}
\end{equation}
where the magnetic field has the standard definition
\begin{equation}
B = {1\over 2} \epsilon_{ij}F_{ij} = {1\over 2} \epsilon_{ij}
(\partial_{i}A_{j}-\partial_{j}A_{i} + ig[A_{i},A_{j}])\label{9}
\end{equation}
We also note that the branch of $\arctan$ is chosen such that
$\Gamma_{\rm eff}^{{\rm PV}, (1)}$ is a continuous function
of $A_{0}$, which reduces, in the zero temperature limit, to
the corresponding CS action.

This is, in fact, the form of the action (but with a non-Abelian
structure because the background is more general) derived 
in \cite{fosco:1997vu} for a much more restrictive gauge
background. In fact, let us now specialize to the case of a vanishing
electric field. In a static background, a vanishing
electric field would correspond to a field configuration satisfying
\begin{equation}
- E_{i} = D_{i}A_{0} = \partial_{i}A_{0} + ig \left[A_{i},A_{0}\right]
  = 0\label{3}
\end{equation}
This would further constrain the relations on the amplitudes
following from the Ward identities in Eq. (\ref{2b}). We would like to
emphasize that the gauge backgrounds in Eq. (\ref{1}) and those in
Eqs. (\ref{2}),(\ref{3}) are inequivalent (although both correspond to
$\vec{E}=0$) in the sense that there is no smooth gauge
transformation which will take one to the other (see appendix {\bf B}
on more details on the consequences of a vanishing electric field at
finite temperature). In this case, it is straightforward to check from
Eq. (\ref{8}) that
\begin{eqnarray*}
\Gamma_{\rm eff}^{{\rm PV}, (\vec{E}=0)} & = & {ig\over 4\pi} \int
d^{2}x\,{\rm
Tr}\,\arctan\left(\tanh {\beta M\over 2} \tan {g\beta
A_{0}(\vec{x})\over 2}\right)\epsilon_{ij}(D_{i}A_{j} -
\partial_{j}A_{i})\\
 & = & {ig\over 8\pi}  \int d^{2}x\ {\rm
Tr}\,\arctan\!\left(\tanh {\beta M\over 2} \tan {g\beta
A_{0}(\vec{x})\over 2}\right)\!\epsilon_{ij}(\partial_{i}A_{j} -
\partial_{j}A_{i})
\end{eqnarray*}
where the first term vanishes upon integration by parts when the
electric field vanishes. This is, in fact, the exact parity violating
effective action that was obtained in \cite{fosco:1997vu} (We note
here that our result, in this limit, differs from that in
\cite{fosco:1997vu} by a
multiplicative factor of ${1\over 2}$. However, we do not fully
understand if this is a real difference, since these authors imply, in
a later publication \cite{cdfosco} that when evaluated in a smooth
manner, there appears a factor of $2$. So, we will ignore this
difference in the multiplicative factor from now on.).  
It is quite surprising that the perturbative calculations with a less
restrictive background seem to yield a parity violating effective
action of exactly the same form. In what follows, we will show that
this  is indeed the  exact parity violating effective action, in the
static limit, when  the electric field vanishes.

Let us consider a fermion interacting with a static background with a
vanishing electric field (see Eqs. (\ref{2}),(\ref{3})). In the
imaginary  time formalism, the action, for such a theory (Fourier
transformed in energy), would have the form
\begin{eqnarray}
S & = & {1\over \beta} \sum_{n} \int d^{2}x\,
\overline{\psi}_{n}\left(i\gamma_{i}D_{i} +
  \gamma_{0}(\omega_{n}- gA_{0}(\vec{x})) - M\right)\psi_{n}\nonumber\\
 & = & {1\over \beta} \sum_{n} \int d^{2}x\,\overline{\psi}_{n}
 \left(i\gamma_{i}D_{i} + \gamma_{0}\tilde{\omega}_{n} -
   M\right)\psi_{n}\label{10}
\end{eqnarray}
Here, we have defined
\begin{equation}
\tilde{\omega}_{n}(\vec{x}) = \omega_{n} - g A_{0}(\vec{x})\label{11}
\end{equation}
which is a nontrivial matrix in the internal space, but is
proportional to the identity matrix in the Dirac space.

As in the Abelian case, following \cite{fosco:1997ei,fosco:1997vu},
let us define
\begin{equation}
\gamma_{0}\tilde{\omega}_{n}(\vec{x}) - M =
\rho_{n}(\vec{x})\,e^{i\gamma_{0}\phi_{n}(\vec{x})}\label{12}
\end{equation}
where
\begin{equation}
\rho_{n}(\vec{x}) = \sqrt{\tilde{\omega}_{n}^{2}(\vec{x}) +
  M^{2}},\quad \phi_{n}(\vec{x}) = \arctan
  {\tilde{\omega}_{n}(\vec{x})\over M}\label{13}
\end{equation}
Here, $\rho_{n},\phi_{n}$ are matrices in the internal space, but are
proportional to the identity matrix in the Dirac space. Furthermore,
since the nontrivial matrix structures for
$\rho_{n}(\vec{x}),\phi_{n}(\vec{x})$ arise  only from their
dependence on $A_{0}(\vec{x})$, it follows (see Eq. (\ref{3})) that
\begin{equation}
\left[\rho_{n}(\vec{x}),\phi_{n}(\vec{x})\right] = 0,\qquad
D_{i}\phi_{n}(\vec{x}) = 0\label{14}
\end{equation}
We note that this is a crucial difference from the derivation in
\cite{fosco:1997vu}.
The special gauge background in \cite{fosco:1997vu}
does satisfy this, but here our derivation is quite general.

In terms of these new variables, the action of
Eq. (\ref{10}) takes the form
\begin{equation}
 S = {1\over \beta} \sum_{n} \int d^{2}x\, \overline{\psi}_{n}
\left(i\gamma_{i}D_{i} +
  \rho_{n}(\vec{x})\,e^{i\gamma_{0}\phi_{n}(\vec{x})}\right)\psi_{n}\label{15}
\end{equation}
If we now make a chiral redefinition (from the point of view of two dimensions)
 of the fermion fields of the form
\begin{equation}
\psi_{n} = e^{-{i\over 2}\gamma_{0}\phi_{n}(\vec{x})}
\tilde{\psi}_{n},\qquad \overline{\psi}_{n} =
\overline{\tilde{\psi}}_{n} e^{-{i\over
    2}\gamma_{0}\phi_{n}(\vec{x})}\label{16}
\end{equation}
it is straightforward to show, using the properties of the matrices
$\rho_{n},\phi_{n}$ discussed above as well as Eq. (\ref{14}), that
the action, Eq. (\ref{15}), takes the form
\begin{equation}
 S = {1\over \beta} \sum_{n} \int d^{2}x\,
\overline{\tilde{\psi}}_{n}\left(i\gamma_{i}D_{i} +
  \rho_{n}(\vec{x})\right)\tilde{\psi}_{n}\label{17}
\end{equation}

From the definitions in Eq. (\ref{13}), it is clear that $\rho_{n}$ is
parity conserving and, therefore, when functionally integrated, the
action in (\ref{17}) will only contribute to the parity conserving part
of the effective action. The parity violating part of the effective
action will arise only from the Jacobian for the field redefinitions
in Eq. (\ref{16})
\[
{\cal D}\overline{\psi}_{n} {\cal D}\psi_{n} = J_{n} {\cal
  D}\overline{\tilde{\psi}}_{n} {\cal D}\tilde{\psi}_{n}
\]
which can be calculated following \cite{fosco:1997vu,gamboa:1981} 
and leads to the parity violating effective action of the form
\begin{eqnarray}
\Gamma_{\rm eff}^{{\rm PV},\,(\vec{E}=0)} & = & \sum_{n} \log
  J_{n} = {ig\over 4\pi} \sum_{n} \int d^{2}x\,{\rm
  Tr}\,\phi_{n}(\vec{x})\epsilon_{ij} F_{ij}\\\label{18}
 & = & {ig\over 8\pi}\,\int d^{2}x\,{\rm Tr}\arctan\!\left(\tanh
  {\beta M\over 2} \tan {g\beta A_{0}(\vec{x})\over
  2}\right)\!\epsilon_{ij}(\partial_{i}A_{j} - \partial_{j}A_{i})\nonumber
\end{eqnarray}
This is, therefore, the exact parity violating effective action in the
static case when the electric field vanishes, independent of the
directions of the gauge potentials in the internal symmetry space. One
can  explicitly check
that this action leads to the parity violating amplitudes calculated
in the static limit (for vanishing electric field) given in
Eqs. (\ref{5})-(\ref{6}).

\section{Manifestly $\vec{E}$ dependent parity violating effective
action:} 

We have also calculated the parity violating amplitudes, up to the
seven point function, in the static limit for the general case when
there is manifest
$\vec{E}$ dependence. For ease of presentation, let us decompose an
arbitrary parity violating amplitude as
\begin{equation}
\Pi_{\mu_{1}\cdots \mu_{n}}^{a_{1}\cdots a_{n}} = \Pi_{\mu_{1}\cdots
  \mu_{n}}^{a_{1}\cdots a_{n}, (1)} +
  \Pi_{\mu_{1}\cdots\mu_{n}}^{a_{1}\cdots a_{n},(2)}\label{19}
\end{equation}
Correspondingly, we will also define
\begin{equation}
\Gamma_{\rm eff}^{\rm PV} = \Gamma_{\rm eff}^{{\rm PV}, (1)} +
\Gamma_{\rm eff}^{{\rm PV}, (2)}\label{20}
\end{equation}
where we identify the second class of terms as manifestly depending on
$\vec{E}$, namely,
\begin{equation}
\Pi_{\mu_{1}\cdots \mu_{n}}^{a_{1}\cdots a_{n}, (2)} =
\Pi_{\mu_{1}\cdots \mu_{n}}^{a_{1}\cdots a_{n}, (\vec{E})},\qquad
\Gamma_{\rm eff}^{{\rm PV}, (2)} = \Gamma_{\rm eff}^{{\rm PV},
(\vec{E})}
\end{equation} 
The amplitudes as well as the effective action for manifestly
$\vec{E}$ independent  terms are
already given in the earlier section. Therefore, in this section, we
will only describe the parts which manifestly depend on $\vec{E}$. Let
us  also define the following notation for any pair of
two-dimensional
vectors, $\vec{a},\vec{b}$, (repeated indices are summed)
\begin{equation}
\vec{a}\cdot \vec{b} = a_{i}b_{i},\qquad \vec{a}\times \vec{b} =
\epsilon_{ij}a_{i}b_{j}\label{21}
\end{equation}

With this, the parity violating amplitudes take the following
forms. Since the color factors are not so complicated for the
amplitudes up to the four point function, we will give their general
forms first,
\begin{eqnarray}
\Pi_{0i}^{ab,\,(\vec{E})} & = &  -
g^{2}\delta^{ab}\epsilon_{ij}k_{j}\left({k^{2}I^{(3)}\over 3} +
  {(k^{2})^{2}I^{(4)}\over 10} + \cdots \right)\nonumber\\
\Pi_{0ij}^{abc,\,(\vec{E})} & = & ig^{3}f^{abc}\left[{I^{(3)}\over
    3}\Big\{\delta_{ij}\vec{k}_{2}\times\vec{k}_{3} - \epsilon_{ij}(2
    k_{2}^{2}+3\vec{k}_{2}\cdot\vec{k}_{3} + 2
    k_{3}^{2})\right.\nonumber\\
 &  & \quad +  (k_{2i}k_{2l} - k_{3i}k_{3l}) \epsilon_{lj} +
   (k_{2j}k_{2l} - k_{3j}k_{3l}) \epsilon_{li}\Big\}\nonumber\\
 &  & \quad -{I^{(4)}\over 20}\left\{\delta_{ij}
   \vec{k}_{2}\times\vec{k}_{3} (3k_{2}^{2} + 2
   \vec{k}_{2}\cdot\vec{k}_{3} + 3 k_{3}^{2})\right.\nonumber\\
 & & \!\!\!\!\!\!\!\!\!\!\!\!\!\!\!\!\!\!\!\!\!\!\!\!\!\!\!\!\!\!  +
 \epsilon_{ij} \left(6\,k_{2}^{4} + 5
   (\vec{k}_{2}\cdot \vec{k}_{3})^{2} + 15 k_{2}^{2}
   \vec{k}_{2}\cdot\vec{k}_{3} + 5 k_{2}^{2}k_{3}^{2} +
   k_{2}\leftrightarrow k_{3}\right)\nonumber\\
 &  & \!\!\!\!\!\!\!\!\!\!\!\!\!\!\!\!\!\!\!\!\!\!\!\!\!\!\!\!\!\! +
 \left(\epsilon_{il}k_{3l}\left(k_{2j}(k_{3}^{2}-k_{2}^{2}) +
     k_{3j}(3 k_{2}^{2} + 4 \vec{k}_{2}\cdot\vec{k}_{3} +
     4k_{3}^{2})\right)-(k_{2}\leftrightarrow k_{3}, i\leftrightarrow
   j)\right)\nonumber\\
 &  &  \!\!\!\!\!\!\!\!\!\!\!\!\!\!\!\!\!\!\!\!\!\!\!\!\!\!\!\!\!\!
\left.\left. - \left(\epsilon_{il}k_{2l}k_{2j} (4k_{2}^{2}
       + 4 \vec{k}_{2}\cdot \vec{k}_{3} + 3 k_{3}^{2}) -
       (k_{2}\leftrightarrow k_{3}, i\leftrightarrow
       j)\right)\right\} + \cdots \right]\nonumber\\
\Pi_{000}^{abc,\,(\vec{E})} & = & - ig^{3}
f^{abc}\,\vec{k}_{1}\times\vec{k}_{2}\,\left(I^{(3)} + {I^{(4)}\over
    2}  (k_{1}^{2} +  \vec{k}_{1}\cdot\vec{k}_{2} + k_{2}^{2}) +
  \cdots\right)\nonumber\\
\Pi_{000i}^{abcd,\,(\vec{E})} & = & {g^{4}I^{(3)}\over 3}
\epsilon_{ij}\Big(  ({2\over N}\delta^{ab}\delta^{cd} + d^{abe}d^{cde}) 
                     (3k_{3}+k_{4})_{j}\nonumber\\
& & \quad\quad\quad\;\; +(b\leftrightarrow c, k_{2}\leftrightarrow
k_{3}) + (a\leftrightarrow c, k_{1}\leftrightarrow k_{3}) +
\cdots\Big) \label{22}\\
\Pi_{0ijl}^{abcd,\,(\vec{E})} & = &
-{g^{4}\over 12} I^{(3)} f^{abe}f^{cde}\left(
  {\cal A}_{0ijl}(k_{2},k_{3},k_{4}) -(j\leftrightarrow l,
  k_{3}\leftrightarrow k_{4}) + \cdots
\right)\nonumber
\end{eqnarray}
where $\cdots$ represent higher order terms in momentum and we have
defined 
\begin{eqnarray}
{\cal A}_{0ijl} (k_{2},k_{3},k_{4}) & = &
\left[\delta_{ij}\epsilon_{lm}(k_{2}+k_{4})_{m} -
  {\delta_{il}\epsilon_{jm}\over 2}
  (k_{2}+2k_{3}+k_{4})_{m}\right.\nonumber \\
 &  & \left.+ \epsilon_{ij} (k_{2}+2k_{3}+k_{4})_{l} +
   {\epsilon_{il}\over 2} (k_{2}-k_{4})_{j} + 
(i\leftrightarrow l, k_{2}\leftrightarrow k_{4})\right]\nonumber
\end{eqnarray}

The higher point functions are simpler to describe for the case of
$SU(2)$ (for simplicity of color factors), where they take the forms
\begin{eqnarray}
\Pi_{00000}^{abcde,\,(\vec{E})} & = &  {ig^{5}\over 2}\left[(3I^{(4)} -
8M^{2}I^{(5)}) \left(\epsilon^{abe}\delta^{cd} \vec{k}_{1}\times
  \vec{k}_{2} + \epsilon^{ace}\delta^{bd} \vec{k}_{1}\times
  \vec{k}_{3} \right.\right.\nonumber\\
 &  & \quad + \epsilon^{ade}\delta^{bc} \vec{k}_{1}\times
  \vec{k}_{4} + \epsilon^{bce}\delta^{ad} \vec{k}_{2}\times
   \vec{k}_{3}\nonumber\\
 &  & \quad \left.\left. + \epsilon^{bde}\delta^{ac} \vec{k}_{2}\times
   \vec{k}_{4} + \epsilon^{cde}\delta^{ab} \vec{k}_{3}\times
   \vec{k}_{4}\right) + \cdots \right]\nonumber\\
\Pi_{000ij}^{abcde,\,(\vec{E})} & = & {2ig^{5}I^{(3)}\over 3} \,\epsilon_{ij}
(\delta^{ab}\epsilon^{cde} + \delta^{ac}\epsilon^{bde} +
\delta^{bc}\epsilon^{ade}) + \cdots\nonumber\\
\Pi_{0ijlm}^{abcde,\,(\vec{E})} & = &\left[{2ig^{5}I^{(3)}\over
    3}\left(\epsilon^{abc}\delta^{de}\epsilon_{ij}\delta_{lm} +
    \epsilon^{abd}\delta^{ce}\epsilon_{il}\delta_{jm} +
    \epsilon^{acd}\delta^{be}\epsilon_{jl}\delta_{im}\right.\right.\nonumber\\
 &  & \quad \left.\left. +
     \epsilon^{abe}\delta^{cd}\epsilon_{im}\delta_{jl} +
     \epsilon^{ace}\delta^{bd}\epsilon_{jm}\delta_{il} +
     \epsilon^{ade}\delta^{bc}\epsilon_{lm}\delta_{ij}\right) + \cdots
 \right]\nonumber\\
\Pi_{00000i}^{abcdef,\,(\vec{E})} & = &
\left[{g^{6}(3I^{(4)}-8M^{2}I^{(5)})\over
    10}\epsilon_{ij}\left((5k_{5}+k_{6})_{j}
    C_{1}^{abcdef}\right.\right.\nonumber\\
 &  & \quad +
    (5k_{4}+k_{6})_{j} C_{2}^{abcdef} + (5k_{3}+k_{6})_{j}
    C_{3}^{abcdef}\nonumber\\
 &  & \quad \left.\left. +
     (5k_{2}+k_{6})_{j} C_{4}^{abcdef} + (5k_{1}+k_{6})_{j}
     C_{5}^{abcdef}\right) + \cdots \right]\nonumber\\
\Pi_{00000ij}^{abcdefg,\,(\vec{E})} & = & -{2ig^{7}\over 5} (3I^{(4)}-
8M^{2}I^{(5)}) C^{abcdefg}\epsilon_{ij} + \cdots\label{24}
\end{eqnarray}
where the $C$'s are defined in Eq. (\ref{7}).

There are several things to note from the structures of these
amplitudes. First of all, we can think of the new structures,
$\Pi_{0i}^{ab,(\vec{E})}$ and $\Pi_{0ij}^{abc,(\vec{E})}$, as higher
order  corrections to
the basic structure in Eq. (\ref{5}). However, structures where all
Lorentz indices are \lq\lq$0$'' or structures with more than two
spatial indices are completely new and are not present in terms that
do not manifestly depend on $\vec{E}$. Furthermore, even the
structures with  one
and two spatial indices in Eqs. (\ref{22})-(\ref{24}) are of the same
order (in powers of momentum) as those in
Eqs. (\ref{5})-(\ref{6}). Thus, unlike in the Abelian case, here the
amplitudes, coming from the parity violating effective action that
does not manifestly depend on $\vec{E}$, cannot be thought of as
leading  order
contributions in the static limit. Furthermore, it can be easily
checked that the amplitudes in Eqs. (\ref{22})-(\ref{24}) satisfy the
Ward identity, Eq. (\ref{2b}), following from the residual gauge
invariance of the static action.

There is one other significant difference between the structure of the
amplitudes in Eqs. (\ref{5})-(\ref{6}) and those in
Eqs. (\ref{22})-(\ref{24}), which also reflects the difference in the
structure of the Abelian and the non-Abelian theories. Namely, it is
easy to check that the only amplitudes in Eqs. (\ref{5})-(\ref{6}) which
survive in the zero temperature limit are the two and the three point
amplitudes. All higher amplitudes in Eqs. (\ref{5})-(\ref{6}) vanish at
zero temperature. In contrast, all the amplitudes in
Eqs. (\ref{22})-(\ref{24}) have a non-vanishing contribution at zero
temperature. Furthermore, we note that all amplitudes with one spatial
index (for example, $\Pi_{000i}^{abcd,\,(\vec{E})}$) are linear in
momentum in the leading order, while those with two spatial indices
(for example, $\Pi_{000ij}^{abcde,\,(\vec{E})}$) have a leading
behavior which is independent of momentum. In general, we note that
all the amplitudes in Eqs. (\ref{22})-(\ref{24}) have a leading momentum
dependence which is of lower order than their Abelian counterpart at
zero temperature \cite{coleman:1985zi}
(The Abelian box amplitude, at zero temperature, for
example, would have a leading momentum dependence that is at least
quartic \cite{Brandt:2000dd}). 
This difference in the behavior of the Abelian and the
non-Abelian amplitudes is a consequence of the nontrivial form of the
Ward identity in the non-Abelian case. In fact, we have explicitly checked
that the amplitudes, Eqs. (\ref{22})-(\ref{24}), do satisfy the
non-Abelian Ward identities at any temperature.

Given the amplitudes in Eqs. (\ref{22})-(\ref{24}), we can also look for
the appropriate action that would give rise to these amplitudes up to
this order. With a lot of work, it can be determined that all
the leading order terms in the amplitudes in Eqs. (\ref{22})-(\ref{24})
can be derived from the effective action
(the normalization is easily determined from the leading order two-point
function in Eq. (\ref{22}))
\begin{equation}
\Gamma_{1,\,{\rm eff}}^{{\rm PV},\,(\vec{E})} = 
-{i g^2 I^{(3)}\over 3}\int
d^{2}x\, {\rm Tr} \left(E_{i}D_{i}B - BD_{i}E_{i} +
  \epsilon_{ij}E_{i}D_{0}E_{j}\right).\label{25}
\end{equation}
In the static limit, of course, the time derivative term in $D_{0}$
gives zero. It is interesting that the relative coefficients between
the $\vec{E}-B$ term and the $\vec{E}-\vec{E}$ term could have been
different, in principle. The fact that they are the same at any
temperature  and have a
nonzero limit at vanishing temperature, suggests that they come from a
single Lorentz invariant structure of the form
\begin{equation}
\Gamma_{1,\,{\rm eff}}^{{\rm PV},\,(\vec{E})} = -{i g^2 I^{(3)}\over 6}
\int d^{2}x\, {\rm Tr}\,\epsilon_{\mu\nu\lambda}
F_{\mu\nu}D_{\alpha}F_{\alpha\lambda}\label{26}
\end{equation}
It is worth pointing out here that there is a second possible Lorentz
invariant structure that is available at this order, namely,
\[
\int d^{2}x\,{\rm
  Tr}\,\epsilon_{\mu\nu\lambda}F_{\mu\alpha}D_{\nu}F_{\alpha\lambda}
\]
which, however, is related to the structure in Eq. (\ref{26}) by
Bianchi identity. Therefore, at this order, the parity violating
effective action has the unique form given in Eq. (\ref{26}).

In going beyond five point amplitudes (and leading order), there are
more possible structures available. We find,
after a lot of analysis, that the rest of the structures in the
amplitudes in Eqs. (\ref{22})-(\ref{24}) including the seven point
functions can be derived from an effective action of the form
\begin{eqnarray}
\Gamma_{2,\,{\rm eff}}^{{\rm PV},\,(\vec{E})} & = & \int d^{2}x\,{\rm
  Tr}\left[c_{1}^{EB}(E_{i}D_{\alpha}D_{\alpha}D_{i}B -
  BD_{\alpha}D_{\alpha}D_{i}E_{i})\right.\nonumber\\
 &  & \quad +
  c_{2}^{EB}(E_{i}D_{\alpha}D_{i}D_{\alpha}B -
  BD_{\alpha}D_{i}D_{\alpha}E_{i})\nonumber\\
 &  & \quad
  \left.+ \epsilon_{ij}\left(c_{1}^{EE}E_{i}D_{\alpha}D_{\alpha}D_{0}E_{j}
  + c_{2}^{EE}
  E_{i}D_{\alpha}D_{0}D_{\alpha}E_{j}\right)\right]\label{27}.
\end{eqnarray}
The four coefficients $c_{1,2}^{EB,EE}$ can all be determined, in principle,
comparing the effective action with the diagrammatic results. For instance,
the two-point function in Eq. (\ref{22}) gives
\begin{eqnarray}\label{coeff1}
c_{1}^{EB} + c_2^{EB} & = & -{ig^2 I^{(4)}\over 10},
\end{eqnarray}
while the $000$ component of the three-point function in Eq. (\ref{22}) as
well as the $00000$ component of the five point function in
Eq. (\ref{24}) yield respectively
\begin{eqnarray}\label{coeff2}
2\,c_{1}^{EE} + c_2^{EE} & = & -{ig^2 I^{(4)}\over 4} \nonumber \\
   c_{1}^{EE} + c_2^{EE} & = & {ig^2\over 20}\left(3 I^{(4)}-8 m^2
   I^{(5)}\right)
\label{28}
\end{eqnarray}
As a consequence of gauge invariance, all the other results in 
Eqs. (\ref{22}) and (\ref{24}) are consistent with the previous
relations (they do not give new relations).
A closed system of equations for all the coefficients would require
further analysis involving the sub-leading contribution to 
the $000i$ component of the four point function. Nonetheless, 
the coefficients $c_{1,2}^{EE}$ are already fully determined
from Eq. (\ref{coeff2}). It is interesting that, 
unlike the earlier case of $\Gamma_{1,\,{\rm eff}}^{{\rm PV},\,(\vec{E})}$, 
here,  $c_{1,2}^{EE}(T)\neq c_{1,2}^{EB}(T)$. 
Therefore, these terms do not combine to a Lorentz
invariant form at finite temperature (Rather, they can be written as
the sum of two terms, one of which is manifestly Lorentz invariant
while the other is not.). However, it is also clear that,
in the limit of zero temperature, when $I^{(5)}={5\over 8 m^2}I^{(4)}$
(see appendix {\bf B}),
Eq. (\ref{coeff1}) and the second relation in Eq. (\ref{coeff2}) are
consistent with $c_{1,2}^{EE}(T=0)= c_{1,2}^{EB}(T=0)$, so that the
zero temperature effective action that is manifestly Lorentz invariant
takes the form
\begin{eqnarray}
\Gamma_{2,\,{\rm eff}}^{{\rm PV},\,(\vec{E})}(T=0) = -{ig^2
  I^{(4)}(T=0)\over 20} 
  \int d^{2}x\,{\rm
  Tr}\,\epsilon_{\mu\nu\lambda}\left(3F_{\mu\nu}D_{\alpha}D_{\alpha}
D_{\beta}F_{ \beta\lambda} \right.\nonumber\\ \left.- 
  F_{\mu\nu}D_{\alpha}D_{\beta}D_{\alpha}F_{\beta\lambda}\right)\label{29}
\end{eqnarray}

Thus, we see that, in the presence of a non-vanishing electric field,
the theory develops a family of parity violating effective actions in
the static limit at finite temperature of the form
\begin{equation}
\Gamma_{\rm eff}^{{\rm PV},\,(\vec{E})} = \Gamma_{1,\,{\rm eff}}^{{\rm
    PV},\,(\vec{E})} + \Gamma_{2,\,{\rm eff}}^{{\rm PV},\,(\vec{E})} +
    \cdots\label{30}
\end{equation}
where the higher order terms can be determined from a calculation of
perturbative amplitudes at higher orders. These actions are manifestly
invariant under the residual gauge transformations in the static
gauge. They have a non-vanishing contribution at zero temperature
completely consistent with the Ward identities of the non-Abelian
theory. In fact, their forms are suggestive and can be trivially
extended to the non-static case in which case they will be invariant
under a general non-Abelian {\it small} gauge transformation.

\section{Going beyond the static limit:}

Our calculations have been strictly in the static limit and have
yielded the static effective action of the form
\begin{equation}
\Gamma_{\rm eff}^{\rm PV} = \Gamma_{\rm eff}^{{\rm PV}, (1)}
+ \Gamma_{\rm eff}^{{\rm PV},\,(\vec{E})}
\end{equation}
with the forms of these actions given in the earlier sections. In
particular, $\Gamma_{\rm eff}^{{\rm PV},\,(\vec{E})}$ contains a family
of terms involving electric and magnetic fields, which can, in
principle, be determined order by order from a calculation of the
amplitudes. These actions are invariant under the residual gauge
transformations in Eq. (\ref{2a}). However, these are {\it small}
gauge transformations and our interest has been to understand the
behavior of the thermal parity violating effective action under a {\it large}
gauge transformation. It is clear that, in the strict static limit,
there can be no {\it large} gauge transformation and we must
necessarily go away from the strict static limit if we want to analyze
the behavior of the effective action under a {\it large} gauge
transformation.

There have been some previous attempts at constructing  thermal {\it
  large} gauge transformations \cite{Pisarski:1987gq,dunne:2001}. For
  example,  in
  \cite{Pisarski:1987gq} , it has been shown,
for the gauge group $SU(2)$, that the gauge transformations
\begin{equation}
U(t,\vec{x},\beta) = \exp [{2\pi i t\over \beta}
\hat{\theta}(\vec{x})\cdot\vec{\sigma} ]
\end{equation}
where $\hat{\theta}(\vec{x})$ is a two dimensional instanton, lead to
a non-vanishing winding number which is even. Although the general
structure of thermal, non-Abelian {\it large} gauge transformations is not yet
fully understood, it is clear that these must be time-dependent
non-Abelian transformations, which are periodic in time with a period
$\beta$. We note from our discussion in the last section that the
structure of the terms in $\Gamma_{\rm eff}^{{\rm PV},\,(\vec{E})}$ is
suggestive and we have, in fact, already written these in a form that
holds for non-static backgrounds. These terms reduce to the
appropriate result in the static limit and are invariant under
non-Abelian gauge transformations, {\it small} and {\it large}. The
main question, therefore, is how to generalize $\Gamma_{\rm eff}^{{\rm
    PV},\,(1)}$ away from the static limit.

As we have argued in the introduction, it is very hard to go beyond
the static limit computationally. For general energies and momenta,
this is rather involved, even in the case of the self-energy
\cite{Brandt:2000dd} and the  level of complexity 
increases enormously as we go to higher point functions in a
non-Abelian background. 
As a result, we have taken an alternate approach. Namely, we have the
effective action in the static limit and we have tried to look for
generalizations, away from the strict static limit, that would reduce
in a natural manner to the static result in the appropriate limit and
will be gauge invariant, only in the simpler case of a vanishing
electric field.

To this end, we take guidance from the earlier
works on Abelian gauge backgrounds and follow as closely as
possible to the structures that arise there, since after all, the
non-Abelian result should yield the Abelian one in the appropriate
limit. In this connection, we recall that, in
an Abelian background with $\vec{E}=0$, the Bianchi
identity implies that the magnetic field is static which allows us to
choose a background of the form
\[
A_{0} = A_{0}(t),\qquad A_{i} = A_{i}(\vec{x})
\]
Subsequently, one can rotate away the time dependence of the $A_{0}$
field, without affecting the time dependence of the $A_{i}$ field, by a
suitable time dependent gauge transformation of the form
\begin{equation}
A_{\mu}\rightarrow A_{\mu} - \partial_{\mu}\Omega,\quad \Omega(t) =
\left(\int_{0}^{t} - {t\over \beta}\int_{0}^{\beta}\right)
dt'\,A_{0}(t')\label{abelian}
\end{equation}
This leads to the time-independent transformed field,
$a_{0}(\vec{x},\beta)$, given by
\begin{equation}
a_{0}(\vec{x},\beta) = {1\over \beta} \int_{0}^{\beta}
dt'\,A_{0}(t',\vec{x})\label{abelian1}
\end{equation}
which becomes space independent for a vanishing electric field, as
mentioned above.

Let us now analyze the corresponding issues in the case of a
non-Abelian background. Here, for a vanishing electric field, Bianchi
identity only implies that
\begin{equation}
\partial_{0}B + ig \left[A_{0},B\right] = D_{0}B = 0\label{31}
\end{equation}
Namely, the magnetic field is not necessarily static, but is
covariantly constant in time. In a
non-static background, in this case, we can again find a time
dependent  gauge transformation which will rotate away the time
dependence in the $A_{0}$ field. Namely, under
\begin{equation}
A_{\mu}\rightarrow \tilde{A}_{\mu} = U^{-1}A_{\mu}U - {i\over g}
U^{-1}\partial_{\mu}U 
\end{equation}
with
\begin{equation}
U(t, \vec{x}) = \left(P e^{-ig\int_{0}^{t}
    dt'\,A_{0}(t',\vec{x})}\right)\left(P e^{ig\int_{0}^{\beta}
    dt'\,A_{0}(t',\vec{x})}\right)^{{t\over\beta}}\label{32}
\end{equation}
where $P$ stands for the path ordering along the time
direction, the time dependence of $A_{0}(t,\vec{x})$ can be
transformed away 
much like in the Abelian case. (We avoid the symbol $T$ for time
ordering so as not to create confusion with temperature.) For
simplicity, we have taken the path ordering with respect to $t=0$,
and assume that $U(t=0,\vec{x}) = 1$.
This, then, leads to the periodicity at finite temperature to
correspond to
\[
U(\beta,\vec{x}) = U(0,\vec{x}) = 1
\]
which can be easily checked to hold true. 

Such a gauge transformation, which preserves the condition (\ref{31}), goes
over to the Abelian one, Eq. (\ref{abelian}), in the appropriate
limit, since, in that case, path ordering is trivial. The
transformation (\ref{32})  also brings
out an interesting feature of the non-Abelian theory, namely, path
ordered quantities \cite{Polyakov:1987ez} do arise in this case in a
natural manner. 

It is worth noting here, for later purposes, that
\begin{equation}
{\rm Tr}\,\left[\left(P e^{-ig\int_{0}^{\beta}
      dt'\,A_{0}(t',\vec{x})}\right)B(0,\vec{x})\right] = {\rm
      Tr}\,\left[\left(P e^{-ig\int_{x_{0}}^{x_{0}+\beta}
      dt'\,A_{0}(t',\vec{x})}\right)B(x_{0},\vec{x})\right]\label{trace}
\end{equation}
which can be verified using the cyclicity of trace, periodicity of
gauge transformations and the fact that $B$ is covariantly
conserved. This relation shows that, although individually the
path ordered exponential and the magnetic field pick up a dependence
on the reference time $x_{0}$ (with respect to which path ordering is
defined), the above combination is, in fact,
independent of the choice of the reference time. Furthermore, both the
path ordered exponential as well as the magnetic field transform
covariantly under a general, periodic non-Abelian gauge
transformation, for example,
\begin{equation}
\left(P e^{-ig\int_{x_{0}}^{x_{0}+\beta} dt'\,A_{0}(t',\vec{x})}\right)
\rightarrow U^{-1}(x_{0},\vec{x})\left(P
  e^{-ig\int_{x_{0}}^{x_{0}+\beta} dt'\,A_{0}(t',\vec{x})}\right)
U(x_{0},\vec{x})\label{transf}
\end{equation}
 
Keeping these properties in mind, let us try to generalize the parity violating
effective action, in the absence of electric fields (see Eq. (\ref{8})
or (\ref{18})), to non-static
backgrounds. To this end, let us note that the transformed field,
$\tilde{A}_{0}(x_{0},\vec{x},\beta)$, which depends on the reference
time $x_{0}$, can be shown to be related to the path ordered
exponential as
\begin{equation}
e^{-ig\beta \tilde{A}_{0}(x_{0},\vec{x},\beta)} = \left(P
  e^{-ig\int_{x_{0}}^{x_{0}+\beta}
  dt'\,A_{0}(t',\vec{x})}\right)\label{relation}
\end{equation}
This implies that, in the static limit, when the time ordering is
irrelevant, $\tilde{A}_{0}(x_{0},\vec{x},\beta)=A_{0}(\vec{x})$, as
expected. Furthermore, in the Abelian case, where the limits of
integration can also be translated, Eq. (\ref{relation}) goes over to
the relation (\ref{abelian1}).

The above observations are quite important since they allow us to write a
generalization of the parity violating static effective action,
Eq. (\ref{8}) (or (\ref{18})) to non-static backgrounds of the form
\begin{eqnarray}
&  & \Gamma_{\rm eff}^{{\rm PV},\,(\vec{E}=0)}  =  
{ig\over 2\pi\beta}
\int_{0}^{\beta} dx_{0}\int d^{2}x\nonumber \\ 
&  & \times\; {\rm Tr} \arctan\left(\tanh
  {\beta M\over 2} \tan {g\beta
    \tilde{A}_{0}(x_{0},\vec{x},\beta)\over 2}\right)
B(x_{0},\vec{x})\label{action}
\end{eqnarray}
We remark here parenthetically that, in view of Eq. (\ref{trace}), the
above integrand is actually independent of $x_{0}$ for vanishing
electric fields. However, for non-vanishing electric fields, the
integrand will depend on the reference point and, in this case,  the
integration over $x_{0}$ is meaningful.

Let us also note that
\begin{equation}
\tan {g\beta \tilde{A}_{0}\over 2} = i\,{1-e^{ig\beta \tilde{A}_{0}}\over
  1+ e^{ig\beta \tilde{A}_{0}}} = i\left[1 - 2
  e^{ig\beta\tilde{A}_{0}} + 2 (e^{ig\beta \tilde{A}_{0}})^{2} + \cdots \right]
\end{equation}
Namely, the tangent can be expanded in terms of powers of the path
ordered exponential. Furthermore, since the path ordered exponential
as well as the magnetic field transform covariantly under a general
non-Abelian gauge transformation, the effective action in
(\ref{action}) will be invariant under {\it small} gauge
transformations. Under a {\it large} gauge transformation, this action
shifts by $\pi n$ (assuming that the magnetic flux is quantized),
where $n$ is an integer depending on the branch of $\arctan$.

Although the effective action in Eq. (\ref{action}) looks superficially similar
to that in Eq. (\ref{8}) (or (\ref{18})), to which it reduces in the
static limit, it has, in fact, a distinct character. It is invariant
under  general periodic non-Abelian gauge transformations and is a
functional of $\tilde{A}_{0}(x_{0},\vec{x},\beta)$, where (see
Eq. (\ref{relation}))
\begin{equation}
\tilde{A}_{0}(x_{0},\vec{x},\beta) =  {i\over g\beta}\,\log \left(P
  e^{-ig\int_{x_{0}}^{x_{0}+\beta} dt'\,A_{0}(t',\vec{x})}\right)
\end{equation}
This is a non-trivial functional of $A_{0}$, in general, and only
in the static limit, does it coincide with $A_{0}(\vec{x})$, as we
have mentioned earlier.

\section{Conclusion:}

In this paper, we have derived the parity violating thermal
effective action induced by $2+1$ dimensional fermions interacting
with a non-Abelian static gauge background from a perturbative
calculation of amplitudes up to the seven point function. We have attempted to
generalize this result to non-static backgrounds in a way that
naturally reduces to the static action as well as the one for Abelian
backgrounds in the appropriate limits, and which is also {\it small}
and {\it large} gauge invariant. The part of the action, which
involves non-vanishing electric fields, $\Gamma_{\rm eff}^{{\rm
    PV},\,(\vec{E})}$ , contains families of terms which are
manifestly gauge invariant. Within each family, there are several
terms, which are related, in a derivative expansion
\cite{Aitchison:1985pp,das:1987yb}, by the non-Abelian
Ward identities. The generalization of $\Gamma_{\rm eff}^{{\rm
    PV},\,(1)}$ in Eq. (\ref{8}) (or (\ref{18})), for the case of
vanishing $\vec{E}$, given by the
action (\ref{action}) seems to be the best that one can do by
following the parallel with the Abelian case as much as is
possible. We conjecture that this may represent a relevant part of the
complete parity violating effective action away from the static
limit, but not the complete action. The reason why this action cannot
represent  the complete answer,
in the absence of electric fields, can be seen as follows. Let us
consider the non-static, induced CS action at zero temperature, which,
then, can be written as
\begin{equation}
\Gamma_{\rm eff}^{{\rm PV},\,(\vec{E}=0)} = {ig^{2}\over 4\pi} \int
d^{3}x\,{\rm Tr}\,\left(A_{0}B - {1\over 2}
  \epsilon_{ij}A_{0}D_{i}A_{j}\right)\label{last}
\end{equation}
Note that the second term in Eq. (\ref{last}) vanishes identically in
the static limit, in view of Eq. (\ref{3}).
Our action (\ref{action}), for $\vec{E}=0$, therefore, represents a
generalization of the first term in Eq. (\ref{last}), which includes
all higher order thermal radiative corrections proportional to the
magnetic field. To obtain the complete parity violating effective
action in this case, one also needs to determine the corresponding
higher order corrections to the second term in Eq. (\ref{last}). This
is a non-trivial open problem that remains to be understood. We would
like to note here that, in the Abelian case, the two terms have, in
fact, the same structure (with different numerical coefficients),
which  explains why the generalization of the
parity violating effective action in the Abelian case is straightforward.

We would like to thank Profs. G. V. Dunne, S. Okubo and J. C. Taylor
for many helpful discussions. This work was supported in part by US DOE Grant
No. DE-FG 02-91ER40685 and by CNPq and FAPESP, Brazil.

\appendix
\section{Explicit low order forms for $I^{(r+1)}$:}
The explicit results for Eq. (\ref{Isum}) can be related to each other 
by differentiation in relation to $M^2$. Using this simple property and
the basic formula
\begin{equation}
I^{(2)}(T) = {MT\over 
8\pi}\sum_{n=-\infty}^{\infty}{1\over\left(M^2+\omega_n^2\right)}=
{1\over 8\pi} \tanh\left({M\over 2T}\right),
\end{equation}
we have derived the following results
\begin{equation}
I^{(3)}(T) = {1\over 32 M T \pi}\left({T\over M}
\tanh\left({M\over 2T}\right)
+{1\over 2}\tanh^2\left({M\over 2 T}\right)-{1\over 2}\right),
\end{equation}
\begin{eqnarray}
I^{(4)}(T) & = &   {1\over 64 M^3 T \pi}\left[
\left({T\over M}-{M\over 6 T}\right)
\tanh\left({M\over 2 T}\right)\right.\nonumber\\
& & \qquad 
\left. + {1\over 2}\tanh^2\left({M\over 2 T}\right)
+{M\over 6 T}\tanh^3\left({M\over 2 T}\right)
-{1\over 2}\right],
\end{eqnarray}
\begin{eqnarray}
I^{(5)}(T) & = & {1\over 512 M^5 T \pi}\left[
\left({5 T\over M}-{M\over T}\right)
\tanh\left({M\over 2 T}\right) \right.\nonumber\\
& & \qquad + \left. \left({5\over 2}-{M^2\over 3 T^2}\right)
\tanh^2\left({M\over 2 T}\right)
+ {M\over T} \tanh^3\left({M\over 2 T}\right) \right.\nonumber\\
& & \qquad + \left. {M^2\over 4 T^2}
\tanh^4\left({M\over 2 T}\right)+
{M^2\over 12 T^2} -{5\over 2} \right]
\end{eqnarray}
and

\begin{eqnarray}
I^{(6)}(T) & = & {1\over 1024 M^7 T \pi}\left[
\left({7 T\over M}-{3M\over 2 T} + {M^3\over 15 T^3}\right)
\tanh\left({M\over 2 T}\right)
\right.\nonumber\\ 
& & \qquad +
\left({7\over 2}-{2 M^2\over 3 T^2}\right)
\tanh^2\left({M\over 2 T}\right) \nonumber\\
& & \qquad +
\left({3M\over 2 T}-{M^3\over 6 T^3}\right)
\tanh^3\left({M\over 2 T}\right) \nonumber\\
& & \qquad +
{M^2\over 2 T^2}\tanh^4\left({M\over 2 T}\right)
+{M^3\over 10 T^3}\tanh^5\left({M\over 2 T}\right)
 \nonumber\\ 
& & \qquad\qquad + 
\left.{M^2\over 6 T^2} -{7\over 2} \right].
\end{eqnarray}

\section{Vanishing electric field at finite temperature:}

In this appendix, we will describe some of the consequences of a
vanishing electric field, both in the Abelian as well as the
non-Abelian theory, which will also clarify why our choice of
backgrounds is inequivalent to those in \cite{fosco:1997vu}.

First, let us consider the Abelian theory where a vanishing electric
field implies
\begin{equation}
E_{i} = \partial_{0}A_{i} - \partial_{i}A_{0} = 0
\end{equation}
The Bianchi identity, in this case, leads to
\begin{equation}
\partial_{0} B = 0,\qquad B = {1\over 2} \epsilon_{ij}F_{ij}
\end{equation}
so that the magnetic field is static and determines
\begin{equation}
A_{i} (\vec{x},t) = \bar{A}_{i}(\vec{x}) + \partial_{i}\alpha
(\vec{x},t)
\end{equation}
The vanishing of the electric field condition, then, determines
\begin{equation}
A_{0}(\vec{x},t) = \partial_{0} \alpha(\vec{x},t)
\end{equation}

At zero temperature, it is clear that we can make a gauge
transformation to set the $A_{0}$ field to zero. For example, since
under a gauge transformation,
\begin{equation}
A_{\mu} \rightarrow A'_{\mu} = A_{\mu} - i U^{-1}\partial_{\mu} U
\end{equation}
we can choose
\begin{equation}
U = e^{-i\Omega}
\end{equation}
where
\begin{equation}
\Omega = \alpha (\vec{x},t) = \int_{0}^{t} dt'\,A_{0}(\vec{x},t')
\end{equation}
which would yield
\begin{equation}
A_{0}(\vec{x},t) \rightarrow 0,\qquad A_{i}(\vec{x},t) \rightarrow
\bar{A}_{i}(\vec{x}) 
\end{equation}
In other words, at zero temperature, in the Abelian theory, the same
gauge transformation that sets $A_{0}$ to zero also makes $A_{i}$
static.

At finite temperature, however, the gauge fields as well as the gauge
transformations have to be periodic. In this case, choosing
\begin{equation}
\Omega(\vec{x},t) = \int_{0}^{t} dt'\,A_{0}(\vec{x},t') - {t\over
\beta} \int_{0}^{\beta} dt\,A_{0}(\vec{x},t)
\end{equation}
we obtain (only true for vanishing electric fields)
\begin{equation}
A_{0}(\vec{x},t)\rightarrow A'_{0}= {1\over \beta} \int_{0}^{\beta}
dt\,A_{0}(\vec{x},t),\qquad A_{i}(\vec{x},t)\rightarrow
 A'_{i} = \bar{A}_{i}(\vec{x})
\end{equation}
Thus, in the Abelian theory, even at finite temperature, the same
gauge transformation makes $A_{0}$ and $A_{i}$ simultaneously
static. However, the scalar potential can no longer be set to
zero. Note also that, although  $A'_{0}$  is seemingly space
dependent, it is in fact a constant when the electric field vanishes,
since 
\begin{equation}
\partial_{i}A'_{0} = {1\over \beta}\int_{0}^{\beta}
dt\,\partial_{i}A_{0} = {1\over \beta} \int_{0}^{\beta}
dt\,\partial_{0}A_{i} = 0
\end{equation}
because of the periodicity of the fields (and the vanishing electric field).

Let us next consider a non-Abelian theory. Here, the fields are
matrices belonging to some representation of the gauge group and the
vanishing electric field condition implies (we will set the coupling
to unity)
\begin{equation}
E_{i} = \partial_{0}A_{i} - \partial_{i}A_{0} + i[A_{0}, A_{i}] = 0
\end{equation}
The Bianchi identity, in this case, would imply
\begin{equation}
D_{0}B = \partial_{0}B + i[A_{0},B] = 0
\end{equation}
where
\begin{equation}
B = {1\over 2} \epsilon_{ij} F_{ij} = {1\over 2}
\epsilon_{ij}(\partial_{i}A_{j}  -
\partial_{j}A_{i} + i[A_{i},A_{j}]) 
\end{equation}
Let us note that, under a non-Abelian gauge transformation,
\begin{equation}
A_{\mu}\rightarrow A'_{\mu} = U^{-1}A_{\mu}U - iU^{-1}\partial_{\mu}U
\end{equation}
whereas the field strengths, such as $B$ and $E_{i}$, transform covariantly.

We see that, in the non-Abelian theory, a vanishing electric field
does not imply that the magnetic field is static, rather it is
covariantly static. However, the solution to the covariantly static
condition gives
\begin{equation}
B(\vec{x},t) = U B(\vec{x},0) U^{-1},\qquad U =
\left(P e^{-i\int_{0}^{t}dt'\,A_{0}(\vec{x},t')}\right)
\end{equation}
where $U$ involves a path ordered exponential signifying the
non-Abelian nature of the fields. At zero temperature, it is clear
that we  can make a gauge transformation with $U$ defined above which
will make $B$ static, namely,
\begin{equation}
B(\vec{x},t)\rightarrow U^{-1}B(\vec{x},t)U = B(\vec{x},0)
\end{equation}
It is also easy to check that, under the same gauge transformation,
\begin{eqnarray}
A_{0}& \rightarrow & A'_{0} = U^{-1}A_{0}U - iU^{-1}\partial_{0}U =
0\nonumber\\
A_{i}& \rightarrow & A'_{i} = U^{-1}A_{i}U - iU^{-1}\partial_{i}U
\end{eqnarray}
where
\begin{equation}
\partial_{0}A'_{i} = 0
\end{equation}
when the electric field vanishes. Namely, at zero temperature, the
same transformation that makes the magnetic field static, also makes
$A'_{i}$ static and $A'_{0} = 0$ for a vanishing electric field. It
trivially follows now that (because $A'_{0}=0$)
\begin{equation}
[A'_{0}, A'_{i}] = 0
\end{equation}

At finite temperature, however, the gauge transformations have to be
periodic. We can generalize the earlier gauge transformation to be
periodic by defining
\begin{equation}
U^{(\beta)} = \left(P e^{-i\int_{0}^{t}dt'\,A_{0}(\vec{x},t')}\right)
e^{it\tilde{A}_{0}(\vec{x})}
\end{equation}
where, periodicity determines
\begin{equation}
e^{-i\beta \tilde{A}_{0}(\vec{x})} = \left(P e^{-i\int_{0}^{\beta}dt\,
A_{0}(\vec{x},t)}\right)
\end{equation}
Under such a gauge transformation,
\begin{equation}
B(\vec{x},t)\rightarrow (U^{(\beta)})^{-1} B(\vec{x},t) U^{(\beta)} =
B(\vec{x},0) 
\end{equation}
which follows from the fact that, at finite temperature, the magnetic
field has to be periodic, which in turn implies that
\begin{equation}
[A_{0}(\vec{x},t), B(\vec{x},0)] = 0
\end{equation}
Namely, even at finite temperature, the magnetic field is static when
the electric field vanishes. In fact, this is a very general feature
at finite temperature, namely, a variable that is periodic in the time
variable and transforms covariantly under a gauge transformation can
be made static, if it is covariantly static.

Under the gauge transformation $U^{(\beta)}$, we have
\begin{eqnarray}
A_{0}(\vec{x},t) & \rightarrow & A'_{0} =
(U^{(\beta)})^{-1}A_{0}U^{(\beta)} - i (U^{(\beta)})^{-1}\partial_{0}
U^{(\beta)} = \tilde{A}_{0}(\vec{x})\nonumber\\
A_{i}(\vec{x},t) & \rightarrow & A'_{i} = (U^{(\beta)})^{-1} A_{i}
U^{(\beta)} - i (U^{(\beta)})^{-1}\partial_{i}U^{(\beta)}
\end{eqnarray}
Note that under this transformation, while  $A'_{0}$ becomes static,
there is no {\em a priori} reason for $A'_{i}$ to be static. Let us
pursue this question a little bit more in detail.

The vanishing electric field condition, in terms of the new fields,
reads (electric field transforms covariantly)
\begin{eqnarray}
D'_{0}A'_{i} & \equiv & \partial_{0}A'_{i} + i[A'_{0},A'_{i}] =
\partial_{i}A'_{0}\nonumber\\
{\rm or,}\qquad \tilde{D}_{0}A'_{i} & = & \partial_{i}\tilde{A}_{0}
\end{eqnarray}
We note that, since $A'_{i}$ is not covariantly static, it is not {\em
a priori} clear that it will be static even when we impose 
periodicity at finite temperature. 

In fact, we can solve the above equation exactly and the general
solution has the form
\begin{equation}
A'_{i}(\vec{x},t) = e^{-it\tilde{A}_{0}}\left[A'_{i}(\vec{x},0) +
\int_{0}^{t}dt'\, e^{it'\tilde{A}_{0}} (\partial_{i}\tilde{A}_{0})
e^{-it'\tilde{A}_{0}} \right] e^{it\tilde{A}_{0}}
\end{equation}
It is clear from this that, if
\begin{equation}
[\tilde{A}_{0}(\vec{x}), A'_{i}(\vec{x},0)] = 0,\qquad {\rm and}\qquad
[\tilde{A}_{0}(\vec{x}),(\partial_{i}\tilde{A}_{0}(\vec{x})] = 0
\end{equation}
then, using periodicity of the fields, we can conclude that $A'_{i}$
is static. Let us recall that we still have the freedom of doing a
time-independent gauge transformation. However, it is hard to imagine
that a single gauge transformation can achieve both these conditions
simultaneously. In fact, let us show next that this cannot be achieved
unless some further condition is imposed.

Let us define
\begin{equation}
[\tilde{A}_{0},(\partial_{i}\tilde{A}_{0})] = M_{i}(\vec{x}),\qquad
[\tilde{A}_{0}, A'_{i}(\vec{x},0)] = N_{i}(\vec{x})
\end{equation}
If we now make a gauge transformation $\bar{U}(\vec{x})$, then,
\begin{equation}
\bar{A}_{0} = \bar{U}^{-1}\tilde{A}_{0}\bar{U},\quad \bar{A}_{i} =
\bar{U}^{-1}A'_{i}\bar{U} - i\bar{U}^{-1}\partial_{i}\bar{U}
\end{equation}
It is straightforward to check that such a transformation can achieve
the first of the conditions provided
\begin{equation}
[[\tilde{A}_{0},(\partial_{i}\bar{U})\bar{U}^{-1}],\tilde{A}_{0}] =
-i M_{i}(\vec{x})
\end{equation}
while, for the second, we need
\begin{equation}
[\tilde{A}_{0}, (\partial_{i}\bar{U})\bar{U}^{-1}] = i N_{i}(\vec{x})
\end{equation}
The two conditions can, then, be shown to be compatible provided
\begin{equation}
[\tilde{A}_{0}, \partial_{i}\tilde{A}_{0} +
i[A'_{i}(\vec{x},0),\tilde{A}_{0}]] = 0
\end{equation}
In general, however, there is no reason why this should
hold and, unlike in the Abelian theory, in a non-Abelian theory at
finite temperature, we cannot go to a static configuration, even if
the electric field vanishes.

However, given a vanishing electric field, we can always choose
specific backgrounds that will solve this. The static background
$A_{0}=A_{0}(\vec{x}),A_{i}=A_{i}(\vec{x})$ with $D_{i}A_{0} = 0$
would solve this, as will the background
$A_{0}=A_{0}(t),A_{i}=A_{i}(\vec{x})$ with $[A_{0},A_{i}]=0$. However,
as we have already seen,
these are {\em background choices} that cannot necessarily be implemented
through a gauge transformation. Furthermore, it is not possible to
transform the first background to the second by a gauge
transformation. This is simply seen by noting that if we have such a
gauge  transformation, $U$, it must necessarily be time dependent and
satisfy
\begin{eqnarray}
\partial_{i}\left(U^{-1}A_{0}U - iU^{-1}\partial_{0}U\right) & = &
0\nonumber\\
\partial_{0}\left(U^{-1}A_{i}U - iU^{-1}\partial_{i}U\right) & = &
0\nonumber\\
\left[U^{-1}A_{0}U - iU^{-1}\partial_{0}U , U^{-1}A_{i}U -
iU^{-1}\partial_{i}U\right] & = & 0
\end{eqnarray}
These are three independent conditions, and as we have already seen
earlier with two conditions, they cannot automatically be satisfied
simultaneously. At least, it cannot be done through well behaved and
smooth gauge transformations, which can be seen as follows.

Note that, with a vanishing electric field, the parity violating part
of the effective action can be written, in the static background, as
\begin{equation}
\Gamma_{\rm eff}^{{\rm PV,}\,(\vec{E}=0)} = {ig\over 4\pi} \int
d^{2}x\, {\rm Tr}\,\arctan\left(\tanh {\beta M\over 2}\tan {g\beta
A_{0}(\vec{x})\over 2}\right)\epsilon_{ij}\partial_{i}A_{j}(\vec{x})
\end{equation}
On the other hand, in the background of \cite{fosco:1997vu}, this
effective action has the form
\begin{equation}
\Gamma_{\rm eff}^{{\rm PV,}\,(\vec{E}=0)} = {ig\over 2\pi} \int
d^{2}x\, {\rm Tr}\,\arctan\left(\tanh {\beta M\over 2}\tan {g(\int dt
A_{0}(t))\over 2}\right)\epsilon_{ij}\partial_{i}A_{j}(\vec{x})
\end{equation}
Without worrying about the difference in the overall multiplicative
factor, we note that the latter expression is a total derivative (since
$A_{0}$ is only a function of $t$) and will vanish
unless the background configurations are singular. This makes it clear
that we cannot go from the static background to the second background
through a smooth and regular gauge transformation.

\section{Abelian nature of PV action in a special gauge background:}
In this appendix, we will point out briefly why the parity violating
effective action, restricted to the particular background field
configuration of Eq. (\ref{1}) cannot have a non-Abelian
structure. Let us note that the particular background in Eq. (\ref{1})
leads to a vanishing electric field and, therefore, we are necessarily
talking about $\Gamma_{\rm eff}^{{\rm PV},\,(\vec{E}=0)}$. As we will
argue, it is the last condition in Eq. (\ref{1}),
\begin{equation}
\left[A_{0}, A_{i}\right] = 0\label{a1}
\end{equation}
which is particularly restrictive and does not allow any non-Abelian
structure in $\Gamma_{\rm eff}^{{\rm PV},(\vec{E}=0)}$. In fact, let
us first show explicitly that the effective action in \cite{fosco:1997vu}
contains no non-Abelian structure. When Eq. (\ref{a1}) holds, it follows that
\begin{equation}
\left[(A_{0})^{n}, A_{i}\right] = 0\label{a'}
\end{equation}
where $n$ is any arbitrary integer. Using this as well as the
cyclicity of trace, it follows trivially that
\begin{equation}
{\rm Tr}\,\left((A_{0})^{n}\left[A_{i},A_{j}\right]\right) = 0
\end{equation}
This shows that the non-Abelian terms in the effective action in
\cite{fosco:1997vu} are, in fact, absent as a consequence of Eq. (\ref{a1}).

Let us now show this in general. We note that Eq. (\ref{a1}) implies
one of the following two possibilities.

i) The obvious solution to Eq. (\ref{a1}) is that $A_{0}$ and $A_{i}$
are parallel in the internal space. In this case, the field
configurations are truly Abelian. In this case, though, let us note
that
\begin{equation}
{\rm Tr}\, A_{0}A_{i} \neq 0\label{a2}
\end{equation}

ii) The second possibility will be to have $A_{0}$ and $A_{i}$
orthogonal in the internal space in a special way. For example, when
considering $SU(N)$, we can always choose a basis such that we have
$SU(N-1)\oplus U(1)$ embedded in $SU(N)$. In this case, we can choose
\begin{equation}
A_{0}\in U(1),\qquad A_{i}\in SU(N-1)\label{a3}
\end{equation}
 and they will satisfy Eq. (\ref{a1}). In fact, if $N$ is sufficiently
 large, we can choose a basis to embed $SU(N-m)\oplus SU(m)\oplus
 U(1), m>1$
 in $SU(N)$. In such a case, we can choose
\begin{equation}
A_{0}\in SU(m)\oplus U(1),\qquad A_{i}\in SU(N-m)\label{a4}
\end{equation}
and they will commute. In this case, the gauge field configurations
will have truly non-Abelian character. However, since $A_{0}$ and
$A_{i}$ belong to orthogonal spaces in this case, we will have
\begin{equation}
{\rm Tr}\, A_{0}A_{i} = 0\label{a5}
\end{equation}

As we have already shown in section {\bf 2} (or as can also be seen in a
derivative expansion \cite{Aitchison:1985pp,das:1987yb}), 
in the absence of electric field, the parity
violating part of the effective action has the form
\begin{equation}
\int d^{2}x\,{\rm Tr}\, F(A_{0}) B\label{a6}
\end{equation}
Such an action will vanish for the second possibility with a truly
non-Abelian nature of the gauge field configurations, while it will be
nonzero only for the first possibility where the gauge field
configurations have an Abelian character. In other words, the last
condition in Eq. (\ref{1}) is too restrictive and necessarily forces
the parity violating part of the effective action, in the absence of
electric fields, to have only an Abelian structure.

\end{document}